\documentclass{osa-article}

\journal{osajournal}

\articletype{Research Article}

\begin{document}

\title{Sinogram super-resolution and denoising convolutional neural network (SRCN) for limited data photoacoustic tomography}

\author{Navchetan Awasthi,\authormark{1} Rohit Pardasani,\authormark{1} Sandeep Kumar Kalva,\authormark{2}
        Manojit Pramanik,\authormark{2} and Phaneendra K. Yalavarthy\authormark{1,*}}

\address{\authormark{1}Department of Computational and Data Sciences, Indian Institute of Science, Bangalore 560012\\
\authormark{2}School of Chemical and Biomedical Engineering, Nanyang Technological University,
 637459, Singapore}

\email{\authormark{*}yalavarthy@iisc.ac.in} 



\begin{abstract}
 The quality of the reconstructed photoacoustic image largely depends on the amount of photoacoustic (PA) boundary data available, which in turn is proportional to the number of detectors employed. In case of limited data (owing to less number of detectors due to cost/instrumentation constraints), the reconstructed 
PA images suffer from artifacts and are often noisy. In this work, for the first time, a deep learning based model was developed to super resolve and denoise the photoacoustic sinogram data. The proposed method was compared with existing nearest neighbor interpolation and wavelet based denoising techniques and was shown to outperform them both in numerical and {\em{in-vivo}} cases. The improvement obtained in Root Mean Square Error (RMSE) and Peak Signal to Noise Ratio (PSNR) for the reconstructed PA image using the sinogram data 
that was super-resolved and denoised using proposed neural network based method  was as high as 41.70 ${\%}$ and 6.93 dB respectively compared to utilizing limited sinogram data.
\end{abstract}


\section{Introduction}
Photoacoustic (PA) tomography offers optical resolution at ultrasonic depth and does not involves introduction of instruments inside the body i.e. non-invasive in nature\cite{xia2014small,wang2012photoacoustic,pramanik2008design,upputuri2016recent,zhou2016tutorial}.  It involves irradiating the region of interest with a nanosecond laser pulse\cite{zhou2016tutorial}. The tissue chromophores absorbs the laser pulse leading to an increased temperature of the tissue locally. This small temperature rise because of the thermoelastic expansion in a pressure wave resulting in the generation of PA wave. These waves propagate through the tissue  under observation and are acquired by the ultrasonic detectors placed at the boundary of the imaging domain\cite{xia2014small,zhou2016tutorial}. These acoustic waves collected  at the boundary are then used to reverse the forward function for estimation of the initial pressure rise distribution which is the inverse problem. 

The photoacoustic tomographic imaging is essentially an initial value problem with the aim of reconstructing the initial value at time ${t=}$ 0 given the pressure information at time ${t}$ at the boundary. Various techniques are present for reconstructing the initial pressure distribution, which includes analytical algorithms (delay-and-sum and backprojection (BP)) and time-reversal (TR) algorithms \cite{wang2012investigation,wang2011imaging,rosenthal2013acoustic,awasthi2018vector,huang2013full,awasthi2018image}. The computational efficiency of these methods comes at the cost of huge data requirement, as the amount of data used for reconstruction governs the quality of PA image in these methods. The liabilities to acquiring large data are increased scan time and expensive experimental setup. Also, the setups used may cover only an aperture for the PA tomographic measurements and may not cover the whole object which results in limited data availability\cite{xu2004reconstructions,arridge2016adjoint,arridge2016accelerated}.

\begin{figure*}[h]
\centering
\includegraphics[width=\linewidth,height = 12 em]{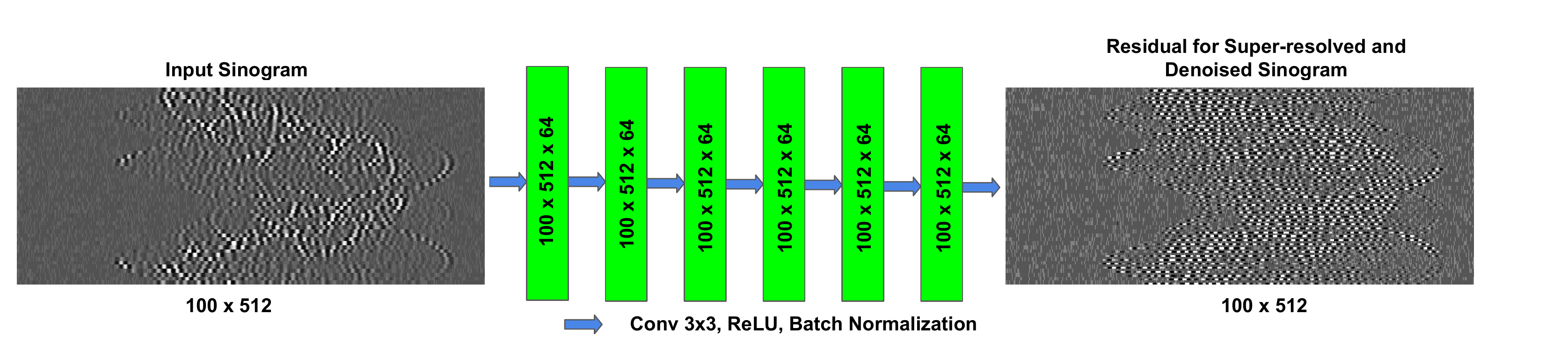}
\caption{The proposed Sinogram super-resolution and denoising convolutional neural network (SRCN) deep neural network architecture with seven convolutional layers. The sinogram which was upsampled by nearest neighbour interpolation is given as input (patches of $S_{IN}(512\times100)$) and the output is the residual map for the denoised sinogram (corresponding patches of $S_R(512\times100)$. Each convolution layer had 64 filters except the last, which had only one filter. Each convolution operation was followed by ReLU activation and batch normalization.}
\label{fig1}
\end{figure*}

The reconstruction using analytical or time-reversal methods results in poor image quality in terms of the contrast especially when the data available is limited. The need of large data can be met by using interpolation based methods to super-resolve and denoise the photoacoustic sinogram. Alternatively, model based techniques were proposed to improve the quantitative accuracy for the PA images reconstructed for limited data cases. These algorithms are robust to noise and are computationally complex compared to analytical algorithms \cite{wang2012investigation,rosenthal2013acoustic,awasthi2018vector,wang2011imaging,buehler2011model,dean2012acceleration,dean2012accurate,awasthi2018image,huang2013full,paltauf2002iterative}. {The $k-$ wave based interpolation\cite{treeby2010k}, which is widely used, typically involves nearest neighbor/linear interpolation on the incomplete/limited data to obtain better image reconstruction. This interpolation main utility is to improve the reconstructed image obtained using time-reversal, i.e. outgoing wave from each discrete detector position in the time-reversal process interacts with other positions on the measurement surface at which a pressure value is also being enforced. This interaction can be improved by interpolation. Even if sufficient detectors are available for capturing the data for an acceptable reconstruction quality, the interpolation may not be able to mitigate the effect of noise present in the sinogram data.}

   {In Ref. \cite{antholzer2018deep}, a Convolutional Neural Network (CNN) was proposed for improving the PA reconstruction obtained using filtered backprojection algorithm. On optimization of data, in Ref. \cite{anas2018enabling},    {the authors proposed a Recurrent Neural Network (RNN) that improves the quality of images and reduces scan time by exploiting temporal information. In similar line of work\cite{anas2018robust}, a CNN was proposed to beamform the channel data to a PA image to improve the reconstruction. In \cite{awasthi2019pa}, a CNN was proposed to fuse the characteristics of different images formed using analytical and model-based inversion methods to get an improved PA image reconstruction.} The above mentioned methods mostly applied in the image space and 
there has been very little work that involved sinogram data. The proposed work here uses only the sinogram domain data and hence was not compared with the above techniques.} 

In general, deep learning has greatly influenced the domain of medical imaging, especially in terms of providing a fully data 
driven model in interpolation of missing data in sparse view sinograms of Positron Emission Tomography (PET) and was shown to perform better than other state of the art methods such as Linear and Directional Interpolation\cite{lee2018deep}. This was inspired by this progress, where a convolutional neural network (CNN) model has been proposed to interpolate and denoise the data to approximate the missing information in the photoacoustic sinogram.    {In \cite{lee2017view}, another network based on U-net architecture was proposed for super resolution of PET sinogram which was not attempted here as PET and PA data characteristics largely differ.}

\section{   {Image Reconstruction in Photoacoustic }}
   {PA image forward model reconstruction involves the computation of the pressure waves acquired by the ultrasonic transducers. The propagation of PA waves is given by the following equation \cite{zhou2016tutorial}
\begin{equation}
\label{pat}
\nabla^2 P(y,t)-\frac{1}{c^2}\frac{\partial^2 P(y,t)}{\partial t^2} = \frac{-\beta}{C_p}\frac{\partial H(y,t)}{\partial t}\,, 
\end{equation}
The notations are defined as:
\begin{itemize}
    \item $H(z,t)$ : the energy deposited per unit time per unit volume
    \item $P(y,t)$ : the pressure at time and position as $t$ and $y$ respectively
    \item $\beta$ : thermal expansion coefficient
    \item $c$ : the sound speed
    \item $C_p$ : specific heat capacity
\end{itemize}
}
\subsection{   {k-Wave Time Reversal Method}}

   {Time Reversal can be performed using K-wave toolbox \cite{treeby2010k} and it is a single step image reconstruction method. Let {\bf T} be the maximum duration for which the PA wave travels inside the imaging domain \cite{hristova2008reconstruction}. It assumes that the solution vanishes outside this time stamp {\bf T}. The initial conditions were assumed to be zero, and the model solves it backward in time to give the initial pressure distribution at time `t'=0. Time reversal is capable of providing a model based resolution
depending on the amount of available data (owing to the number of detectors) as well as the bandwidth of the ultrasonic detectors \cite{huang2013full,hristova2008reconstruction}. In cases, the boundary data acquired by the transducers is limited, the interpolated data (obtained using $k-$wave toolbox) is utilized for estimating the initial pressure distribution \cite{treeby2010k}. This technique can be applied for reconstruction of the initial pressure distribution for full bandwidth data as well as limited bandwidth data.} 

\subsection{{Automated Wavelet Denoising of Recorded Photoacoustic Data}}

   {Various techniques are available for denoising \cite{holan2008automated} the data before doing the PA reconstruction. One of the technique is performing wavelet denoising using maximum overlap DWT (MODWT) \cite{percival2006wavelet}. The main advantages of MODWT (non-orthogonal transform) are:
   \begin{itemize}
       \item The zero padding is not required as the sample size is not defined only for powers of 2.
       \item It applies a filter having zero phase which results in lining the original signal with the features.
   \end{itemize}
    A complete overview of maximum overlap DWT (MODWT) can be obtained in Ref. \cite{percival2006wavelet} and for applications in PA imaging please see Ref. \cite{holan2008automated}. In this work, MODWT was utilized to implement the wavelet smoothing and denoise the interpolated PA noisy signals. The threshold is set automatically using the universal threshold criteria \cite{donoho1994ideal}. }

\subsection{   {Sinogram super-resolution and denoising convolutional neural network (SRCN) (Proposed)}}

   {Convolutional Neural Networks (CNNs) are gaining lot of importance as they are being utilized in various tasks in image processing and reconstruction \cite{girshick2014rich,krizhevsky2012imagenet}. CNNs comprise of activation layers, pooling layers, convolutional layers and batch normalization layers. In this work, we utilized a seven layer CNN with convolutional layers, activation and batch normalization  layers. Each convolutional layer convolves 3*3 size kernels while moving one pixel at a time. Each convolutional layer has 64 filters except the last layer which has only one channel. The structure of CNN used is given in Fig. \ref{fig1}. 
   The loss function used for training of the proposed SRCN network was Mean-squared error (MSE). For better performance, residual learning technique \cite{kim2016accurate} is used to train the network. Residual training involves training the network to predict the difference between the input and the ground truth. This has been shown to give better convergence as compared to training with ground truth as target. The MSE can be written as
\begin{equation}
    MSE = \frac{1}{N} \sum_{i=1}^{N}{\| \hat{y}_i-\phi(x_i) \|^2}
\end{equation}
where ${\hat{y}_i}$ denotes the expected residual of the network while ${\phi(x_i)}$ denotes the output of the CNN.
}

\begin{figure*}[h]
\centering
    \includegraphics[width=\linewidth,height = 18.5 em]{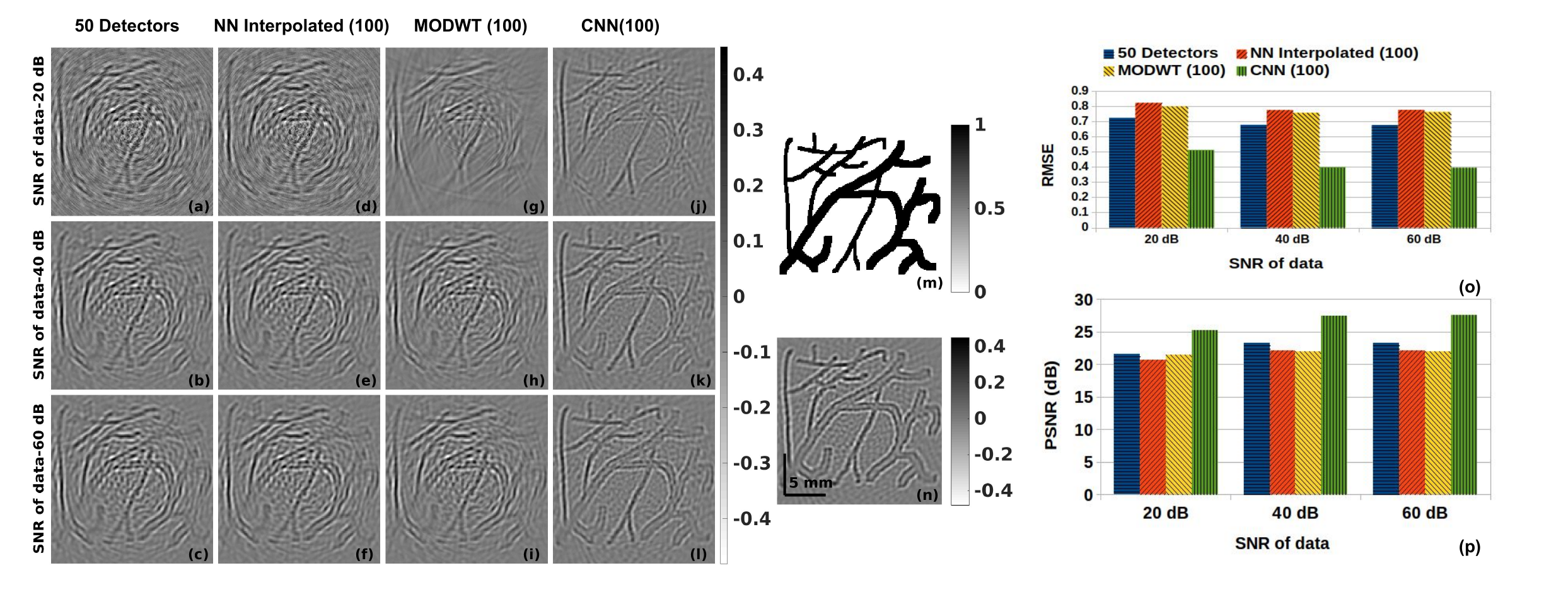}
\caption{Comparison of reconstructed PA images of numerical blood vessel phantom using sinogram data obtained 
using different methods discussed in this work with SNR of sinogram being listed 
against each row (20, 40, and 60 dB). The target blood vessel phantom is shown in (m), while the reconstruction obtained using 100 detectors without noise is shown in (n) to serve as ground truth in calculating RMSE and PSNR. 
The reconstructed PA image using sinogram data of (a-c) 50 detectors (d-f) 100 detectors that were nearest neighbour interpolated (g-i) 100 detectors interpolated and denoised using MODWT method (h-l) 100 detectors super resolved and denoised using proposed CNN based method. The figures of merit, RMSE and PSNR, for the reconstructed 
results were given in (o) ad (p) respectively with (n) serving as ground truth.}
\label{fig2}
\end{figure*}

\begin{figure*}[h!]
\centering
    \includegraphics[width=\linewidth,height = 18.5 em]{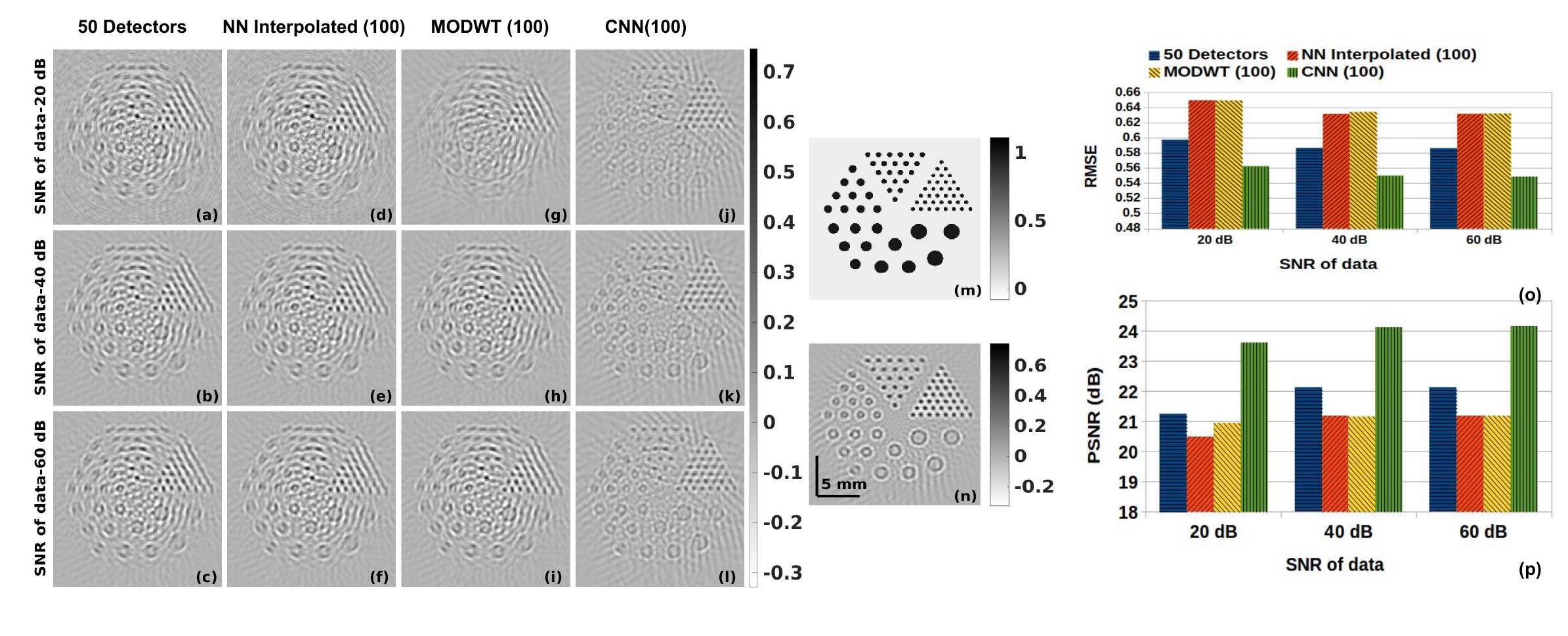}
\caption{Comparison of reconstructed PA images of numerical Derenzo phantom using sinogram data obtained 
using different methods discussed in this work with SNR of sinogram being listed 
against each row (20, 40, and 60 dB). The target derenzo phantom is shown in (m), while the reconstruction obtained using 100 detectors without noise is shown in (n) to serve as ground truth in calculating RMSE and PSNR. 
The reconstructed PA image using sinogram data of (a-c) 50 detectors (d-f) 100 detectors that were nearest neighbour interpolated (g-i) 100 detectors interpolated and denoised using MODWT method (h-l) 100 detectors super resolved and denoised using proposed CNN based method. The figures of merit, RMSE and PSNR, for the reconstructed 
results were given in (o) ad (p) respectively with (n) serving as ground truth.}
\label{fig3}
\end{figure*}

\begin{figure*}[h]
\centering
    \includegraphics[width=\linewidth,height = 18.5 em]{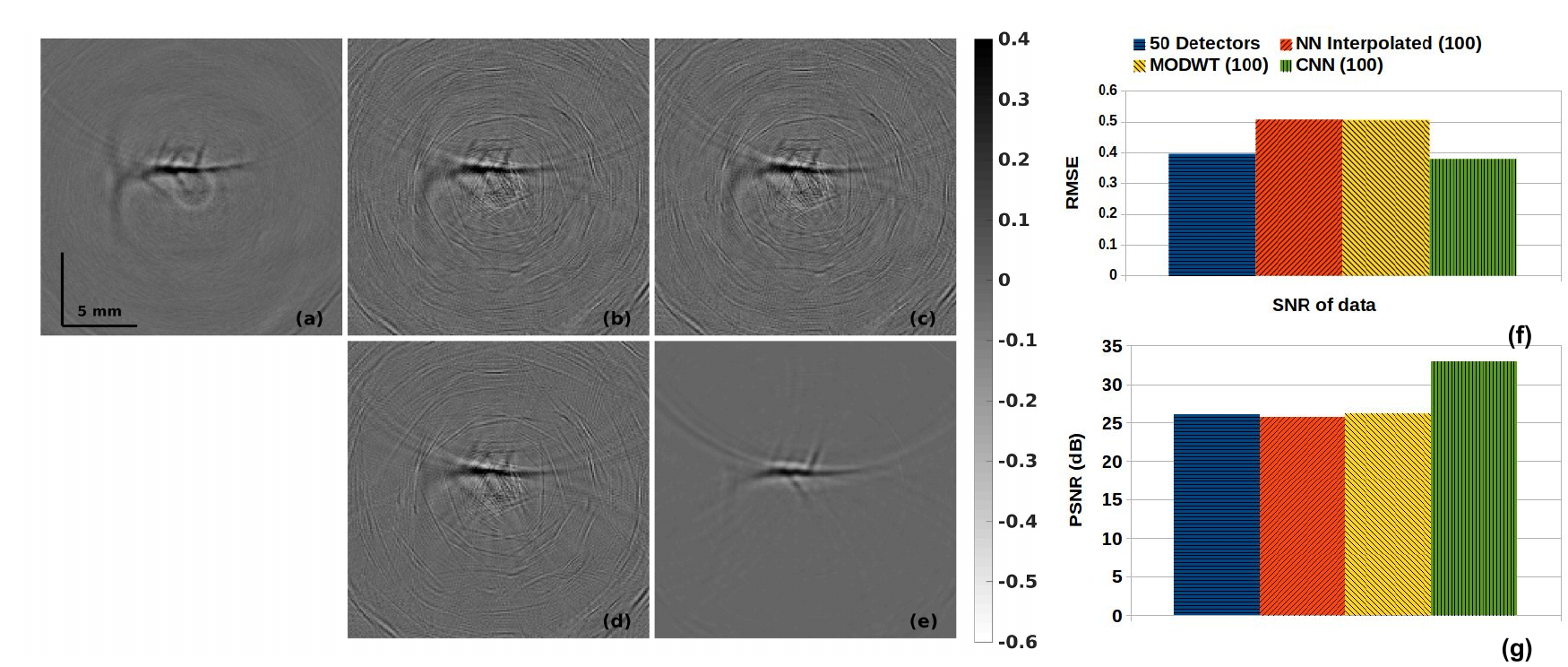}
\caption{Comparison of reconstructed PA images of \textit{in vivo} rat brain sinogram data obtained 
using different methods discussed in this work. 
The reconstructed rat brain PA image using original 100 detectors data is shown (a) to serve as ground truth that is achievable. Reconstruction result using 50 detectors data is shown (b). 
The reconstructed result using 100 detectors sinogram data obtained using nearest neighbor interpolated is given in (c), MODWT method in (d), the proposed CNN method result is shown in (e). RMSE and PSNR (in dB) for these results are shown in (f) and (g), respectively with (a) being taken as ground truth.}
\label{fig4}
\end{figure*}

For training the deep learning model, the sinogram($S$) with 100 detectors was utilized and the steps that generated the patches can be summarized by the following pseudo-code.\\

Step-1: $S_{F} = Noise\ Free\ Sinogram\ with\ 100\ detectors\\
(512 \times 100)$ generated using $k-wave$ on high dimensional grid.\\ 
Step-2: $S_{F\ (512 \times 100)}\xrightarrow[(SNR\ 20/\ 40/\ 60\ dB)]{add\ Gaussian\  Noise\ } S_{FN\ (512 \times 100)}$\\
Step-3: $S_{FN\ (512 \times 100)}\xrightarrow[factor\ of\ two]{sub-sample\ by} S_{HN\ (512 \times 50)}$\\
Step-4: $S_{HN\ (512 \times 50)}\xrightarrow[Interpolation]{Nearest\ Neighbor} S_{IN\ (512 \times 100)}$\\
Step-5:$Residual:S_{R(512 \times 100)} = S_{IN(512 \times 100)} - S_{F(512 \times 100)}$\\

The CNN gets trained on patches of $S_{IN}$ as input and $S_{R}$ as expected output. The proposed SRCN architecture is given in Fig. \ref{fig1}.

The dataset consists of 4600 images extracted from the databases CHASE\cite{fraz2012ensemble}, DRIVE\cite{staal2004ridge}, and STARE\cite{hoover2000locating}, to result in 4600 sinograms for training and validation of proposed model. Fifty random patches of size $32\times32$ were extracted from each of these sinograms, resulting in 230,000 sinogram patches. Out of these, 200,000 patches were used for training and the remaining 30,000 were used for validating the network. The network (given in Fig. \ref{fig1}) consists of seven convolutional layers with each layer followed by a rectified linear unit (ReLU) and batch normalization. The network architecture is similar to the 
one used in Ref. [\citenum{lee2018deep}], with reduced number of layers (20 to 7). {The total number of trainable parameters were 186,497 while the number of non-trainable parameters were 640 (total number of parameters in the network to 187,137).} The rationale behind such reduction is that the problem of PAT sinogram super-resolution has lesser complexity owing to sinogram representing a smooth function (ray sum), which can be easily learnable with reduced number of layers. The initialization was performed using random normal distribution having standard deviation as 0.001 and mean as 0.0 for all kernel weights and all biases were initialized with zeros. The loss function and the optimizer used were mean square error and Adam for training \cite{kingma6980method}. The learning rate was set to be $2\times10^{-8}$ while the first momentum and the second momentum were set to be 0.9, and 0.999 respectively. All computations, including training, were performed on computer having Dual Intel Skylake Xeon 4116 (24 cores) with a clock speed of 2.10 GHz with 64GB RAM consisting of two NVIDIA Tesla P100 12GB GPU cards. The batch size was chosen to be 100 and the model was trained for approximately 11.5 days. Around 11,000 epochs were ran in this duration as each epoch run time was about 91 seconds. Keras\cite{chollet2015keras} using Tensorflow\cite{abadi2016tensorflow} as the backend was used for writing the code for testing and training the network. Once the CNN model was trained, the model was utilized in following manner with $T$ representing the Testing sinogram:\\

\indent
$T_{HN} = Noisy\ Sinogram\ with\ 50\ detectors\ data$\\
\indent
$T_{HN\ (512 \times 50)}\xrightarrow[Interpolation]{Nearest\ Neighbour}T_{IN\ (512 \times 100)}$\\
\indent
$T_{IN\ (512 \times 100)}\xrightarrow{CNN} T_{R\ (512 \times 100)}$\\
\indent
$T_{F\ (512 \times 100)} = T_{IN\ (512 \times 100)} - T_{R\ (512 \times 100)}$\\

Here $T_{F}$ represents the predicted noiseless interpolated sinogram by the trained CNN, which becomes the input to the PA image reconstruction algorithm (in here, time-reversal). 

\section{   {Figures of Merit}}
   {For comparing the efficiency of the proposed methods, the following figures of merit were used for numerical simulations and experimental datasets.}
\subsection{{Root Mean Square Error (RMSE)}}
   {It is an absolute metric to compare the reconstruction quality and is defined as \cite{gandhi2017photoacoustic,pai2018accuracy} :
\begin{equation}
    RMSE(x^{target},x^{recon}) = \sqrt{\frac{\sum{(x^{target}-x^{recon})^2}}{M}}
\end{equation}
Here, the reconstructed pressure distribution is denoted as ${x^{recon}}$ while the target pressure distribution as ${x^{target}}$, and the total number of pixels by $M$. The lower the value of RMSE, the better is the reconstructed image quality.
}
\subsection{{Peak Signal to Noise Ratio (PSNR)}}
   {It is defined as \cite{arridge2016accelerated}:
\begin{equation}
    PSNR = 10\hspace{0.5 em}log_{10} \Bigg(\frac{(Peak Value)^2}{MSE}\Bigg)
\end{equation}
Here, Peak Value denotes the maximum possible value in the image and MSE denotes the Mean Square Error. The higher the value of PSNR, the better the reconstructed image quality.}

\section{Numerical and Experimental Studies}
The imaging domain that was utilized in this work had dimension of 501${\times}$501. Each pixel is 0.1 mm wide and  thus the size of the imaging domain is 50.1 mm $\times$ 50.1 mm. Hundred detectors are placed equidistantly on a circle of radius of 22 mm in the initial experimental setup. A high dimensional grid of size 401 ${\times}$ 401 was used to generate the sinogram data. To avoid inverse crime a lower dimension grid having size 201 ${\times}$ 201 was used to perform the reconstruction imitating the real experimental scenario. The numerical phantoms have a dimension of 201 $\times$ 201, thus having a size of 20.1 mm ${\times}$ 20.1 mm. The generated data from the high dimensional grid was added with white Gaussian noise to result in signal-to-noise ratio (SNR) levels of 20, 40, and 60 dB. An open source MATLAB toolbox $k-$wave \cite{treeby2010k} was used for generating the data in MATLAB. The number of time samples were 512 and the sampling frequency was chosen as 20 MHz. The acoustic detectors that were used had 2.25 MHz as the center frequency with bandwidth of 70${\%}$. 

Numerical blood vessel and Derenzo phantoms were utilized in here to compare the reconstruction accuracy of the proposed SRCN method and compare it with other techniques. These phantoms were unipolar in nature having `1' for the object of interest and `0' for the background with an initial pressure rise distribution of 1 kPa. An \textit{in-vivo} experimental data from rat brain was also utilized to validate the proposed SRCN deep neural network architecture. The data acquisition setup details and the experimental setup details are available in Ref. \cite{jiang2017broadband}. {Note that all the animal experiments conducted here as part of the work followed the regulations and guidelines accepted by the institutional Animal Care and Use committee of Nanyang Technological University, Singapore (Animal Protocol Number ARF-SBS/NIE-A0263).}

\section{Results and Discussion}

The image reconstruction obtained using the proposed CNN based interpolated sinogram data ($T_F$) was compared with the reconstruction obtained using original fifty detectors ($T_{HN}$) and the reconstruction obtained using the nearest neighbour interpolated sinogram for hundred detectors ($T_{IN}$). 
To prove the efficacy of CNN based interpolation and denoising, an automated denoising of the recorded photoacoustic data using Maximum Overlap Discrete Wavelet Transform (MODWT) \cite{holan2008automated} was also utilized and the reconstructed results were compared with the proposed SRCN deep neural network architecture. {The denoising threshold was automatically chosen using universal threshold criteria. 
The reconstructed results were compared quantitatively using RMSE and PSNR with ground truth being 
reconstructed result obtained using original 100 detectors data ($S_F$).} 

The reconstruction PA images using the sinogram obtained using the methods discussed till now, including proposed CNN, were presented in Fig. \ref{fig2} for the blood vessel phantom for SNR of sinogram being 20, 40, and 60 dB (arranged row wise respectively). The target phantom is shown in Fig. \ref{fig2}(m) and 
ground truth reconstruction using 100 detectors data (noise free) is shown in Fig. \ref{fig2}(n). The results obtained using 50 detectors sinogram data are shown in the first column (a-c). Reconstruction results for the interpolated sinogram for 100 detectors are shown in second column (d-f). The 
reconstructed PA images obtained using denoised sinogram data obtained via MODWT method are shown in the third column (g-i). The reconstructed PA images using the proposed CNN based interpolated denoising method were presented in fourth column (j-l). The reconstructed 
images using sinogram obtained via the proposed deep learning based method
was capable of mitigating aliasing artifacts and improvement in visibility of edges. The reconstructed PA image background is less noisy for the result 
obtained using proposed method. The RMSE and PSNR (in dB scale) of reconstruction results are shown in Figs. \ref{fig2}(o) and \ref{fig2}(p) respectively. The improvement (decrease) in RMSE obtained for 20 dB case (first row) was 29.14${\%}$ while for 40 (Second row) and 60 (third row) dB cases, the improvement it achieved was 41.16${\%}$, and 41.70${\%}$ respectively compared to results obtained to using only 50 detectors data. Similarly the improvement in PSNR obtained for 20, 40, and 60 dB cases was 3.68, 4.21, and 4.26 dB respectively.
The same trend was also observed for the case of numerical Derenzo phantom (Fig. \ref{fig3}). In this case, the small radius circles were indistinguishable
due to the aliasing artifacts in all results except proposed CNN based method. 
The improvement for the sinogram data having SNR of 20 dB in terms of RMSE and PSNR was 5.87${\%}$ and 3.14 dB respectively. For sinogram SNR being 40 dB, the improvement was 6.25${\%}$ and 2.95 dB respectively and for sinogram SNR being 60 dB, 6.28${\%}$ and 2.95 dB 
improvement in RMSE and PSNR was observed respectively using proposed method in comparison to other standard 
methods (near neighbour interpolated and MODWT based denoised methods).

An \textit{in-vivo} rat brain sinogram data was also utilized in this work to show the superiority of the proposed SRCN  deep  neural  network  architecture in pre-clinical imaging. The reconstruction result using original hundred detectors sinogram data is shown in Fig. \ref{fig4}(a), which serves as ground truth. The reconstruction using fifty detectors data is shown in Fig. \ref{fig4}(b) and the reconstruction obtained using the data after interpolation is shown in Fig. \ref{fig4}(c). The 
reconstructed PA image obtained using denoised sinogram data utilizing MODWT method was shown in Fig. \ref{fig4}(d). The reconstructed PA image using sinogram 
data of 100 detectors that were denoised and super resolved using the proposed deep learning based method was shown in Fig. \ref{fig4}(e). From these results, it is clear that the proposed CNN based method was able to 
provide super-resolved and denoised sinogram that enabled good quality 
PA images with reduction in artifacts and background noise. The RMSE and PSNR improvements were shown in Fig. \ref{fig4}(f) and \ref{fig4}(g) respectively. The PSNR improvement obtained in the proposed method as compared to the fifty detectors data was 6.93 dB.    {Since the in-vivo data was not acquired in the same fashion as the other numerical phantoms data, there are some structures missing in the denoised improved output, but the overall PSNR improved with reduction in artifacts with the proposed method. Since interpolation depends on the sampling of the data, the change in configuration leads to different set of data collected by the ultrasonic transducers. If the same configuration is used for training as well as testing as was shown for the numerical test phantoms, no loss of structures are observed with improvement in image quality.} 
{Even though this work shown only super-resolving and denoising of sinogram data from 50 detectors to 100 detectors with 512 time samples (making it limited data case), the proposed approach is in general applicable to super-resolving the sinogram of any dimension. Python and MATLAB codes for dataset creation, model construction, training and testing network were provided as an open source \cite{awasthi2019sinogram} to help users to replicate the proposed approach. Note that the training time could be further reduced with usage of multi-GPU framework (current set-up utilized only two GPUs).}

\section{Conclusion}
The proposed sinogram super-resolution and denoising method that is deep learning based clearly shows significant improvement (PSNR improvement as high as 6.93 dB) in the reconstructed PA image. The absence of noise in super-resolved sinogram results in reduced artifacts and better reconstructed image quality. The results presented here were unseen by the proposed deep learning network and  demonstrate that the proposed network was capable of providing a generalizable 
model for super-resolving and denoising the sinogram. As deep learning is making strides in medical imaging, the proposed work clearly demonstrates the 
utility of the same in terms of improving limited experimental data in terms 
of super-resolving and denoising. The proposed approach was fully data driven and authoritatively shown to improve the reconstructed PA image quality using both numerical and {\em{in-vivo}} cases. 

\section*{Disclosures}

The authors declare that there are no conflicts of interest related to this article.

\bibliography{sample}

\end{document}